\documentstyle[aps,pre,psfig]{revtex}
\begin{document}
\bibliographystyle{prsty}
\def\IC{{{{\rm {\mbox{\small l}}} \kern -.50em {\rm C}}}}
\def\IR{{{{\rm l} \kern -.15em {\rm R}}}}
\def\IN{{{{\rm l} \kern -.15em {\rm N}}}}
\def\IZ{{{{\rm Z} \kern -.35em {\rm Z}}}}
\title{Moving lattice kinks and pulses: an inverse method}
\author
{S. Flach and  Y. Zolotaryuk}
\address{
Max-Planck-Institut f\"{u}r Physik komplexer Systeme, 
N\"othnitzer Str. 38, 01187 Dresden, Germany }
\author{K. Kladko} 
\address{Condensed Matter Physics Group and Center for 
Nonlinear Studies,\\
Theoretical Division Los Alamos National Laboratory, \\
MS-B258, Los Alamos, New Mexico 87545, USA}
\date{\today}

\maketitle


\begin{abstract}
We develop a general mapping from given kink or pulse shaped 
travelling-wave
solutions including their velocity to the equations of motion on 
one-dimensional lattices which support these solutions.  
We apply this mapping - by definition an inverse method -
to acoustic solitons in chains with nonlinear intersite interactions,
to  nonlinear Klein-Gordon chains, to reaction-diffusion equations
and to discrete nonlinear Schr\"odinger systems. 
Potential functions can be found in at least a unique way provided
the pulse shape is reflection symmetric and pulse and kink shapes
are at least $C^2$ functions.
For kinks we discuss the relation of our results to the problem
of a Peierls-Nabarro potential and continuous symmetries.
We then generalize our method to higher dimensional lattices
for reaction-diffusion systems. We find that increasing also the number
of components easily allows for moving solutions.
\end{abstract}

\widetext
%
\section{Introduction}

Finding exact travelling-wave (TW) solutions of  
nonlinear lattice systems has been a problem of growing interest
in recent years.
Apart from some integrable systems which support 
TW solutions (e.g., \cite{mt89}), a little is
known about non-integrable discrete systems. 
It appears to be difficult to prove the existence of such  waves
because one has to deal with differential equations with advance 
and delay terms (on some general properties of these equations see
\cite{jkhsmvl93}).
The existence of acoustic (pulse) solitary waves as travelling-wave solutions 
in lattices with nonlinear intersite interactions has been 
proved in \cite{gfjadw94}. However, no proof
is available for other types of solitary waves, for instance,
topological
solitons in nonlinear Klein-Gordon (KG) lattices or other 
discrete kink-bearing systems (an exception is
given in \cite{sjo78}).
Stationary breathers have been shown to be generic solutions
for lattice systems (for a review and further references see
\cite{sfcrw98}). Again the question of whether moving breathers
on lattices exist is still not answered, although a number
of approaches to the subject are known 
\cite{ht92-jpsj-1},\cite{ckks93},\cite{fw94},\cite{cat96},
\cite{satc98},\cite{sfkk98}).
An exception is the case of the 
integrable Ablowitz-Ladik equation \cite{al76}. 

Here we approach the TW existence problem from the inverse side -
we show that for a given TW profile, corresponding
equations of motion can be
generated, so that these  equations of motion yield the chosen  
TW profile as a solution. This has been done first in 
\cite{vhs79} for the $\tanh$-shaped kink and 
extended to reaction-diffusion-type systems in \cite{pcbgr97}.
However, in both cases the analysis was performed for a
specific class of profiles, whereas we will approach this
problem from a general point of view. This in turn will allow us to
obtain general information about the properties of the TW solutions.

The structure of the paper is as follows. In Section II we introduce
the equations of motion. Section III
is devoted to solutions of the nonlinear Klein-Gordon equation
and of  reaction-diffusion-type systems. In Section IV we study 
chains with nonlinear intersite interactions which admit
acoustic (pulse) soliton solutions. We will refer to this type of
lattices as to {\it acoustic} chains. In Section V we deal
with discrete nonlinear Schr\"{o}dinger-type (DNLS) equation.
Section VI is devoted to the structural stability of solitary
waves, and Section VII generalizes our method to higher space
dimensions. Conclusions are given in Section VIII.

%
%
\section { The equations of motion}

We consider a one-dimensional chain with lattice spacing equal
to unity, 
which describes a system of interacting
particles of unity mass. Such a system has a direct physical
meaning and can describe, for example, simple quasi-one-dimensional
molecular crystals. The interparticle interaction 
potential $ W_{n-n'}(r)$ and the on-site potential 
$V(u)$ are, in general, nonlinear functions: 
\begin{equation}
{\ddot u}_n= -V'(u_n) + \sum_{m}W_{n-m}'(u_{n}-u_m)
\label{2}
\end{equation}
where $u_n$ is the displacement of the $n$th particle
from its equilibrium position and $m,n$ are integers.
If the second derivative $\ddot u$ in Eq. (\ref{2}) is replaced
by the first derivative $\dot u$, we obtain a system of 
reaction-diffusion equations.

Another system of interest is 
a generalized discrete nonlinear Schr\"{o}dinger equation (DNLS) 
\begin{eqnarray}
\nonumber
i {\dot \phi}_n &+& C(\phi_{n+1}-2 \phi_n+\phi_{n-1})\\
&+&F(|\phi_n|^2)(\phi_{n-1}+\phi_{n+1})+ G(|\phi_n|^2)\phi_n=0~
\label{3}
\end{eqnarray}
which appears in various fields. 
Here $\phi_n(t)$ is a complex-valued function and $F$ and $G$ 
are general nonlinear functions. 

We are not aware of any systematic approach which shows the
existence or even obtains analytical expressions for TW solutions
of the equations from above.
Therefore we approach the problem
from the opposite side. We formulate an inverse method of creating
the potentials $V$ or $W$ or the pair of functions $(F,G)$ for a
given TW solution.

\section{Solutions of the nonlinear Klein-Gordon equation} 

For sake of simplicity 
let us consider the case of harmonic intersite interaction 
$W_{n-n'}(r)=(Cr^2/2)\delta_{n,(n'\pm 1)} $.
Then the equation of motion (\ref{2}) becomes the well-known nonlinear
Klein-Gordon equation:
\begin{equation}
\ddot u_n=C (u_{n+1}-2u_n+u_{n-1})-V'(u_n).
\label{4}
\end{equation}
First, let us study the dispersion law for small-amplitude waves 
oscillating around some minimum $u_{min}$ of the potential $V(u)$. After
linearising the on-site potential around the above-mentioned minimum
the dispersion law can be written as follows
\begin{equation}
\omega_q^2=\gamma + 2C(1-\cos {q})
\label{4a}
\end{equation}
where $q$ is a wave number and $\gamma=V''(u_{min})$. The group velocity
$s_0=\partial \omega_q /\partial q$ attains its maximal value $s_{max}$ when
$\partial s_0 /\partial q=0$:
\begin{equation}
s_{max}=\sqrt{C- \frac{\sqrt{\gamma^2 +4 \gamma C}-\gamma}{2}}~.
\label{4b}
\end{equation}
We are interested in TW solutions, i.e., solutions that
 propagate with a permanent shape and velocity:
\begin{equation}
u_n(t)=u(n-st) \equiv u(z)~,~~~ z=n-st~,
\label{5}
\end{equation}
where $s$ is the velocity of the travelling wave. As a result, we obtain 
a differential equation with delay and advance terms:
\small
\begin{equation}
s^2 u''(z)=C[u(z+1)-2u(z)+u(z-1)]-V'[u(z)]\;\;.
\label{6}
\end{equation}
\normalsize

\subsection{Moving pulses}

First we consider solutions of a bell-shaped localized 
form, i.e., pulses. 
Given the profile $u(z)$ and its velocity $s$,
we can generate the on-site potential $V$. The function $u(z)$
should satisfy the following conditions:
\begin{itemize}
\item $u(z \rightarrow \pm \infty) \rightarrow 0 $ 
\item $u(-z)=u(z)$
\item $u(z)$ is monotonic in $[0,+\infty [ $
\item $u(z)$ is analytic in 
$ [ 0,+\infty [ $~~~. 
\end{itemize}
To show that the potential $V$ can be generated in a unique way,
we rewrite Eq. (\ref{6}) in the following form:
\begin{eqnarray}
\nonumber V'[u(z)] &\equiv& f[u(z)]=-s^2 u''(z)\\
&+&C[u(z+1)-2 u(z) +u(z-1)]\;\;.
\label{7}
\end{eqnarray}
Now we see that there exists a unique correspondence between
the function $u(z)$ and the force function $f(u)$. Since
$u(z)$ is analytic, we can always rewrite $u(z \pm 1)$ in terms of
$u(z)$. Therefore for each $u(z)$ with the conditions listed
above the function $f(u)$ can be
uniquely defined. The potential $V$ is then obtained by integrating
$f(u)$ once.
This result does not change if we consider a more complicated
interaction potential $W$ which incorporates anharmonic terms
and long range interactions. Thus, we showed here, that for a
given interaction potential $W$, a given pulse profile $u(z)$ which
satisfies
the above conditions and a given velocity of the pulse
we always generate a unique on-site potential $V$ which supports
this TW profile as an exact solution of the equations of motion.

What would happen if we loose the symmetry condition on $u(z)$?
Consider two values $z_1,z_2$ at which $u(z_1)=u(z_2)$. 
Consequently,
the argument of the force function is also the same. But the rhs of
Eq. (\ref{7}) will be different for $z_1$ and $z_2$ in general, which
implies that we obtain two different values for the force function
at the same argument - a circumstance impossible for standard functions.
Thus, we have to require the symmetry of $u(z)$ which guarantees that
the force function is defined in a unique way. This is also the reason
why we can exclude the existence of 
more complicated pulse forms like anti-symmetric
pulses, symmetric pulses with several maxima etc.

Let us investigate some properties of $V$. Since for $z\rightarrow 
\infty$ $u \rightarrow 0$ and $u'' \rightarrow 0$, we find
that $f(0)=0$ which was to be expected. To get more information
about the dependence of $f(u)$ for small $u$ (which tells us
about the stability of the TW solution) we need the leading order
dependence of $u$ on $z$ for large $z$ which is given by the
ansatz of the TW profile. 
Let us assume that our ansatz yields an exponential decay
of $u(z)$ for large distances, i.e.,
\begin{equation}
u(z \rightarrow \pm \infty) \sim e^{-\mu |z|}~,~~ \mu >0~.
\label{8}
\end{equation}
After substituting
Eq. (\ref{8}) into Eq. (\ref{7}) we obtain
\begin{equation}
f(u) \simeq - u[\mu^2 s^2 - 2C(\cosh \mu-1)]~.
\label{9}
\end{equation}
The slope of the  force $f(u)$ for small $u$
changes its sign when $C$ crosses the  value $C_1$ given by 
\begin{equation}
C_1=\frac{s^2 \mu^2} {2(\cosh \mu-1)}.
\label{10}
\end{equation}
This means that the potential $V(u)$ has maxima at $u=0$
when $C<C_1$ and minima if $C>C_1$. 
Consequently, the asymptotic state $u(z \rightarrow \pm \infty)=0$ 
is a dynamically unstable one
for $C < C_1$ and a stable one for $ C > C_1$. 
There is another critical value [see again Eq. (\ref{7})] of 
$C$ given by 
\begin{equation}
C_2=\frac{s^2 u''(0)}{2[u(1)-u(0)]}~.
\label{11}
\end{equation}
If $C > C_2$, an additional extremum (maximum) in $V(u)$ 
appears between $u=0$ and $u(0)$.  
The possible scenarios $C_1< C_2$ and $C_1 > C_2$ are shown in 
Fig. 1(a) and Fig. 1(b), respectively. 
The above statements about 
stability hold for any exponentially decaying pulse. If
the decay is non-exponential, the on-site potential
can become  non-analytic at $u=0$. For example, for a Gaussian tail
\begin{equation}
u(z\rightarrow \pm \infty) \sim e^{-\mu z^2} 
\end{equation}
the on-site potential for small $u$ gets dressed with logarithmic
corrections, e.g. for $\mu=1$
\begin{eqnarray}
\nonumber f(u)=V'(u)&\simeq& u\left[-s^2(2+4 \ln u) \right. \\
 &+&  \left. 2C - \frac{2C}{e}
\cosh \left(2\sqrt{-\ln {u}} \right ) \right]~.
\end{eqnarray}
\normalsize
\begin{figure}
\centerline{
\psfig{figure=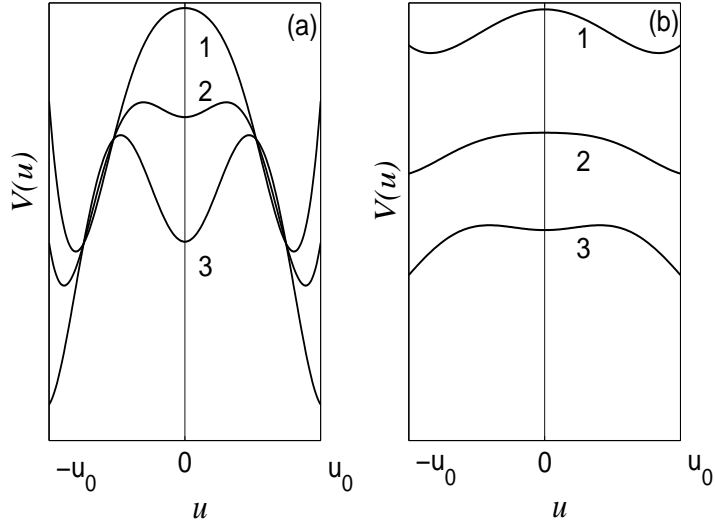,height=200pt,width=270pt,angle=0}
}
\caption[Fig. 1]
{ 
The potential $V(u)$ for a fixed pulse solution: (a) for
$C_1<C_2$ and 1: $C < C_1 < C_2$, 2: $C_1 < C < C_2$,
3: $C_1 < C_2 < C$;  (b) for $C_1>C_2$ and 1: $C < C_2 < C_1$,
2: $C_2 < C < C_1$, 3: $C_2 < C_1 < C$.
The potential is obtained for positive $u$ values and continued
to negative values for sake of transparency.
}
\label{fig1}
\end{figure}
Note that the solution also exists in the ``anti-continuum'' 
limit $C=0$. This seems to be surprising, since the oscillators
are not interacting with each other. Still in this 
case we have a simple equation
\begin{equation}
f[u(z)]=s^2 u''(z)~.
\label{12}
\end{equation}
It can be easily noticed from the bell-shaped form of $u(z)$
that the function $f(u)$ is anti-symmetric and has two zeroes,
one of which is at $u=0$. Thus,
the potential $V(u)$ has a maximum at $u=0$ and a minimum at some
value $0 < u < u(0)$
(see Fig. 2). The separatrix trajectory
corresponds to the motion of each particle from 
the maximum of the potential $V(u)$ to the right wall and back.
This motion needs infinite time.
\begin{figure}[htbp]
\centerline{
\psfig{figure=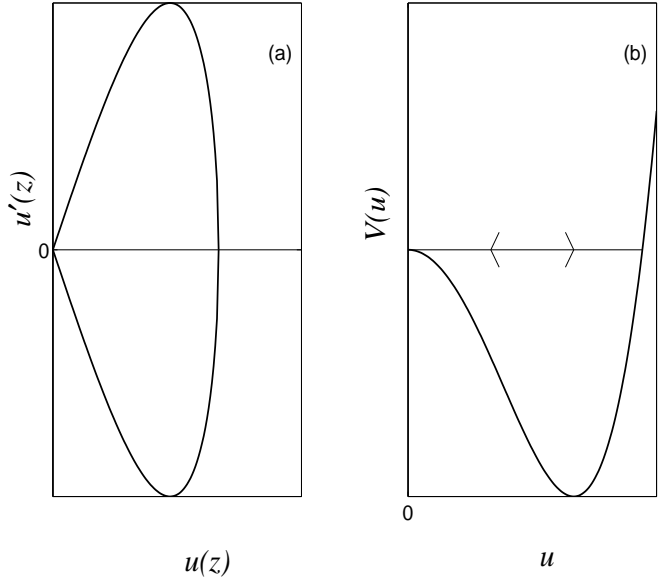,height=220pt,width=250pt,angle=0}
}
\caption[Fig. 2]
{ 
The schematic representation of the pulse solution
in the anti-continuum limit.
}
\label{fig2}
\end{figure}

It follows that it is possible to prepare the initial phases
of all particles on this separatrix such that their uncorrelated
motion resembles the motion of a pulse solution through the system.
This solution is dynamically unstable because $V''(0)<0$.

Finally, we consider possible velocities for exponentially 
decaying pulses (\ref{8}).
We want to check whether our pulses can be subsonic ($s<s_{max}$) or 
supersonic ($s<s_{smax}$).
Taking into account that [see Eq. (\ref{9})] 
\begin{equation}
\gamma=C(\cosh \mu-1)-\frac{\mu^2 s^2}{2}~,
\label{12d}
\end{equation}
we compare $s$ with $s_{max}$. Defining $s_0=s_{max}=s$
we find that the pulse can be both subsonic and supersonic since
in this case 
\begin{equation}
s_0^2=C\frac{\cosh \mu +1-\sqrt{(\cosh \mu+1)^2-2(2+\mu^2)}}{2+\mu^2}~
\label{12e}
\end{equation}
Consequently for fixed $C$ and $\mu$, $s$ has to be small enough to satisfy
$C>C_1$ [see (\ref{10})], when $s(C_1)>s_0$. Thus for $s_0<s<s(C_1)$
the solutions are supersonic, while for $s<s_0$ they are subsonic. 

Let us consider an explicit example of the sech-type profile.
Suppose the profile is described by the function
\begin{equation}
u(z)=u_0 \mbox{sech} (\mu z)
\label{13}
\end{equation}
where $u_0$ is the amplitude of the pulse and $\mu$ its inverse width.
Using the expressions 
\begin{eqnarray}
\nonumber
u''(z)&=&u_0 \mu^2[\mbox{sech} (\mu z)- 2\mbox{sech}^3(\mu z)]\\
&=&\mu^2 \left[ u(z)-2 \frac{u^3(z)}{u_0^2}\right] , \\
\nonumber
u(z-1)+u(z+1)&=&2 u_0\frac{\cosh (\mu z)}{\sinh^2 \mu+\cosh^2 (\mu z)}\\
&=&2 u_0^2 \cosh \mu \frac{u(z)}{u_0^2+ \sinh^2 \mu ~ u^2(z)},
\end{eqnarray}
we reconstruct the on-site potential $V(u)$:
\begin{eqnarray}
\nonumber
V(u)&=&-\frac{1}{2}[ s^2 \mu^2 +2C] u^2+
\frac{s^2 \mu^2}{2u_0^2}u^4\\
&+& Cu_0^2 \frac{\cosh \mu}{\sinh^2 \mu}
\ln \left [ \frac{u_0^2}{\sinh^2 \mu}+u^2\right ]~.
\label{16}
\end{eqnarray}
It is easy to check that its shape will change with $C$ exactly as
described above. We can rewrite this potential in the following form:
\begin{equation}
V(u)=-\frac{1}{2}\kappa_2 u^2+
\frac{\kappa_2-2C}{2u_0^2} u^4+C\sqrt{a^2+u_0^2}\ln \left (a^2+u^2\right )
\label{17}
\end{equation}
where the parameters of the solution $s$ and $\mu$ are given by
\begin{equation}
\mu=\mbox{arcsinh} (u_0/a)~,
~s=\frac{\sqrt{\kappa_2-2C}}{\mbox{arcsinh} (u_0/a)}~.
\label{18}
\end{equation}
After proper rescaling constants $C$ and $u_0$ can be
eliminated and, consequently, we can reduce the number
of system parameters to two. However, we are not able
to rewrite the potential $V(u)$ as a function 
independent from the parameters of the solution (\ref{13}):
$s, u_0, \mu$. This means that we cannot answer so far
whether the solution (\ref{13}) comes as a family
of solutions of (\ref{17}) or is a unique solution
of the obtained equations of motion.

\subsection{Moving kinks}

Kinks or topological solitons are solutions which connect 
two minima of the on-site potential $V(u)$. If $V(u)$ has
several equivalent minima, a countable infinite set of
stationary (time-independent) 
kink solutions of the Klein-Gordon equation (\ref{4}) exists
(in contrast to the space-continuous case, where the continuum
groups of translation symmetry provides a smooth family of
stationary kink solutions). Some of these solutions will be local
minima of the total energy, and some will correspond to saddles.
The question whether moving kinks
as TW solutions (\ref{5}) exist is still
open. Some results (see, e.g., \cite{mpmdk84}) suggest
that kinks in discrete lattices experience a so-called 
Peierls-Nabarro barrier. One interpretation of this barrier 
is that it is the energy difference between stable stationary
kinks and unstable stationary kinks. Indeed it is clear, that
to unpin a stable stationary kink, one needs at least this
amount of energy. Another more sophisticated approach - 
the collective coordinate approach - is a projection technique
which aims at accounting for the dynamics of a kink-like object
boosted to move along the lattice.  
By coupling the kinks center of mass coordinates to phonons,
one arrives at the result that a moving kink will radiate, loose
kinetic energy, and finally be trapped (pinned) by the lattice.
Here the barrier appears as a height of maxima in a potential 
which is used to describe the kink motion.

Nevertheless, the analytical result of \cite{vhs79} suggests that
it is possible to construct  an exact moving kink solution.
This result can be generalised for any profile
$u(z)$ that satisfies the following conditions:
\begin{itemize}
\item $u(z \rightarrow \pm \infty)\rightarrow \pm u_0$~
\item $u(z)$ is monotonic in $]-\infty, +\infty[$
\item $u(z)$ is analytic in $]-\infty, +\infty[$~.
\end{itemize}
Contrary to the case of pulse solutions, we do not need to  
require the function $u(z)$ to have certain symmetries, so that we can 
restrict ourselves 
to monotonicity only. A non-symmetric $u(z)$ profile will simply imply
a non-symmetric function $V(u)$. If the above-mentioned 
conditions are satisfied,
we again can uniquely map the function $u(z)$ onto the potential
$V(u)$. For simplicity, we renormalise the variable $u(z)$
by the kink width $u(z) \rightarrow u(z)/u_0$.

If  
\begin{equation}
u( z\rightarrow \pm \infty) \sim \pm \left(1 -  e^{-\mu |z|}\right)~,
\end{equation}
we can perform the asymptotic analysis  
similar to the case of the pulse solution.
If $C<C_1$ [where $C_1$ comes from Eq. (\ref{10})] it follows
$V''(\pm 1)<0$. Therefore the extremal points $u=\pm 1$ are maxima and, 
consequently, the asymptotic groundstates $ u =\pm 1$ are unstable. 
The sign of $V''(\pm 1)$
changes at $C=C_1$ so that for $C>C_1$ these groundstates are stable.
Another critical $C_2$ value is given by 
\begin{equation}
C_2=\frac{s^2 u'''(0)}{2[u'(1)-u'(0)]}~.
\end{equation}
If $C<C_2$ the state $u=0$ is a minimum and if $C>C_2$ it
is a maximum and we have the standard double-well
potential. For details see Fig. 3.

\begin{figure}[htbp]
\centerline{
\psfig{figure=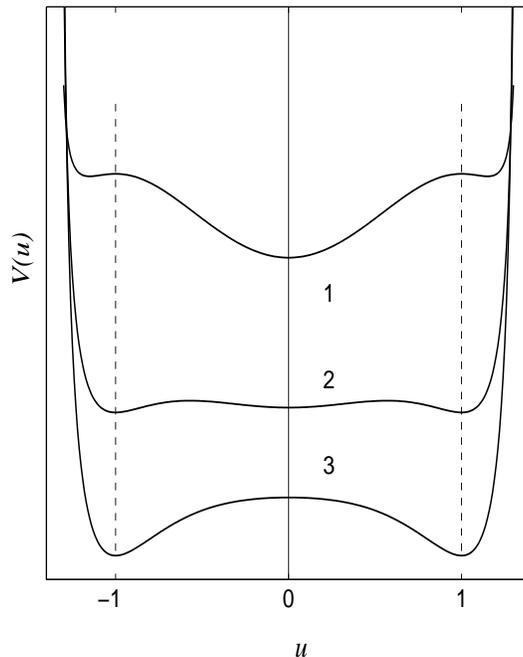,height=250pt,width=200pt,angle=0}
}
\caption[Fig. 3]
{ 
The potential $V(u)$ for a kink solution: 
1: $C < C_1 < C_2$, 2: $C_1 < C < C_2$, and
3: $C_1 < C_2 < C$;  
The potential is obtained for positive $u$ values and continued
to negative values for sake of transparency.
}
\label{fig3}
\end{figure}

For 
\begin{equation}
u(z)=  \tanh (\mu z )
\label{20}
\end{equation}
the explicit form of the potential $V(u)$
was obtained in \cite{vhs79} and \cite{pcbgr97} and 
can be written as follows
\begin{eqnarray}
\nonumber
V(u)&=&(s^2 \mu^2-C)u^2-\frac{s^2\mu^2}{2}u^4\\
&-&\frac{C}{\sinh^2 \mu}\ln (\cosh^2 \mu-u^2 \sinh^2 \mu)~.
\label{21}
\end{eqnarray}
The existence of such a solution does not imply the 
absence of Peierls-Nabarro barrier for stationary kink solutions. 
We demonstrate this 
for the particular case of the solution (\ref{20}). 
We consider a chain with the on-site potential (\ref{21})
with parameters $C=1$, $s=0.5$, $\mu=1$.
Then we calculate two stationary kink solutions, one of which
is stable, and the other one being a saddle in the energy landscape.
Their energies are $E_1 =0.991796$ and $E_2=0.992153$, respectively
(measured relatively to the absolute energy minimum).
The nonzero difference is the Peierls-Nabarro barrier. 
In fact, the existence of a Peierls-Nabarro barrier already follows
from the stability of one of the stationary kink solutions, which 
implies that a finite amount of energy is needed to get out of
the minimum.

Furthermore, we now are in possession of a very effective method
to generate equations of motion which support stationary kinks
{\sl without} a Peierls-Nabarro potential, i.e., where a stationary
kink exists which can be placed anywhere in the chain. Such a system
has a degenerated ground state under the constraint of existence
of one kink. It can be easily generated by putting
$s=0$ in the above inverse method, and that is also true for
pulse solutions. Indeed, in the limit $s \rightarrow 0$ we
generate systems which support kinks (or pulses) which move with
infinitesimally slow velocity. As the system is energy conserving
and the kinetic energy becomes negligible in this limit, the 
groundstate becomes nearly degenerate, and a Goldstone mode appears
for $s=0$. We tested these predictions numerically and obtained
excellent agreement. A particular example of a potential supporting
kinks with zero Peierls-Nabarro potential is 
\begin{equation}
V(u)=-Cu^2 -\frac{C}{\sinh ^2 \mu}\ln (\cosh ^2 \mu -
u^2 \sinh ^2 \mu) \;\;, \label{s1}
\end{equation}
which was obtained for a kink solution $u(z)=\tanh (\mu z)$
from (\ref{21}) by putting $s=0$. It is interesting to note
that Ward and Speight \cite{sw94} also proposed a scheme
which generates systems supporting kink solutions with 
zero Peierls-Nabarro
potential. This scheme uses Bogomolnyi's inequality \cite{ebb76}.
The structure of the potential functions was fixed, but the difference
operators were chosen in an appropriate way. This method then
generates equations of motion which are rather hard to justify
physically. This is not the case when using our inverse method scheme.

\subsection{Reaction-diffusion systems}

Let us also consider 
dissipative systems 
described by the discrete analogue of reaction-diffusion equation
of the form
\begin{equation}
\dot{u}_n=C(u_{n+1}-2u_n+u_{n-1})+f(u_n)~.
\end{equation}
The physical background of this
equation differs from the above considered nonlinear Klein-Gordon
chains but it also
admits localized travelling-wave solutions. These systems are
dissipative because we have a  first order time derivative instead
of the inertia term. Here the function $f(u)$ can have different
meanings, for example, an ion current for nerve fibers 
(see \cite{acs75}).

These systems lack time reversibility. The equation for TW
 solutions  reads
\small 
\begin{equation}
-su'(z)=C[u(z+1)-2u(z)+u(z-1)]+f[u(z)]~. 
\label{s2}
\end{equation}
\normalsize
To generate a system for a moving pulse, let us consider
a symmetric pulse with one maximum as described in the
pulse section of the Klein-Gordon chains. Due to the first
order derivative in Eq. (\ref{s2}) the lhs is anti-symmetric, while
the rhs is symmetric. Consequently, we cannot define $f(u)$ in
a unique way. Any further complication of the symmetry of the
pulse will not help either. Thus, we conclude that there exist
no moving pulse solutions in Eq. (\ref{s2}).

However, reaction-diffusion systems (\ref{s2}) 
support kink solutions. The non-uniqueness problems disappear
as long as the kink shape $u(z)$ is a monotonous function.
Due to the first order derivative in Eq. (\ref{s2}) a kink shape
moving to the right with some given velocity will generate
a function $f_r$ different from the function $f_l$ generated
by the same kink moving with the same velocity but to the left.

As an example, let us consider a profile $u(z)=\tanh (\mu z)$.
Performing the above-mentioned computations we obtain
\begin{equation}
f(u)=-s\mu (1-u^2) - 2C \frac{u}{1+(1-u^2)\sinh^2 \mu}+2Cu
\end{equation}
This result coincides with result of Bressloff and Rowlands 
(see \cite{pcbgr97}).

Finally, let us show that we can also obtain moving pulses provided
we increase the number of components per site. Indeed consider
\begin{eqnarray}
\dot{u}_n&=&C(u_{n+1}-2u_n+u_{n-1})+f(u_n,v_n)\;\;,
\label{s20-1} \\
\nonumber
\dot{v}_n&=&C(v_{n+1}-2v_n+v_{n-1})+g(u_n,v_n)\;\;.
\label{s20-2}
\end{eqnarray}
Assuming $u_n(t)=u(n-st)\;\;,\;\;v_n(t)=v(n-st)$ we find
\begin{eqnarray}
-su'(z)=C[u(z+1)-2u(z)+u(z-1)]+f[u(z),v(z)]\;\;, 
\label{s21-1} \\
-sv'(z)=C[v(z+1)-2v(z)+v(z-1)]+g[u(z),v(z)]\;\;.
\label{s21-2}
\end{eqnarray}
Let us choose a certain profile for $u(z)$. Fixing a value of $u=\kappa$
we obtain a countable set of points $z^{(u)}_i$ such that $u(z_i)=\kappa$.
Here $i$ is an integer and counts all points.
This defines a countable set of functions $z^{(u)}_i(u)$. Similarly
we proceed with $v(z)$. In order to solve the inverse problem, i.e.,
for given functions $u(z)$ and $v(z)$ we have only to require
\begin{equation}
v(z^{(u)}_i) \neq v(z^{(u)}_j)~~ \mbox{if}~~ i\neq j
\;\;,\;\;
u(z^{(v)}_i) \neq u(z^{(v)}_j)~~ \mbox{if}~~ i\neq j\;\;.
\end{equation} 
This is a weak condition satisfied by most choices of $u$ and $v$.
For instance we can even choose symmetric  functions  
having just one maximum and decaying to zero at infinities.
The only restriction would be to shift the centers of the two
functions apart, e.g., $v(z)=u(z-z_0)$ for a given $u(z)$.
But also asymmetric functions with even several maxima are allowed.
Also possible is a symmetric function for $u(z)$ having one maximum
and decaying at infinities, and an asymmetric function for $v(z)$
having the same other properties - with the maxima of both functions
coinciding ($z_0=0$). It is a tedious work to calculate examples,
so in most cases it will be appropriate to obtain the functions
$f$ and $g$ numerically. The reason for the easy construction of
two-component moving pulses is that we introduce two functions 
$f,g$ of two variables, but determine them only on a line in their
phase space $\{u,v \}$. That means that we do not completely define
these functions.

Adding a third component to the problem clearly further relaxes
the conditions on the pulse forms. 
The existence  of two component pulses is partially known  
for space continuous systems
\cite{bkasm94}. Note that our inverse method works as well in the
space continuous case, i.e., where differences of the form
$u_{n+1}-2u_n+u_{n-1}$ are replaced by second derivatives.

Why not doing the same for conservative systems? Then the functions
$f,g$ will be the components of the gradient of some generating
function (e.g., a potential). Thus, we have to impose this
gradient condition, which will restrict the choice of functions
$u,v$.

%
%
\section{ Acoustic chains}

Now let us study systems which support acoustic (pulse) solitary
waves. In these systems the on-site potential is absent
[$V(u)=0$] and the solitary waves appear due to the nonlinearity of
the interaction potential $W(r)$. 
First of all, we introduce the relative displacements, $r_n=u_{n+1}-u_n$.
In these terms the equations of motion take the form
\begin{equation}
\ddot r_n=W'(r_{n+1})-2W'(r_n)+W'(r_{n-1}).
\end{equation}
For TW solutions $r_n(t)=r(n-st)= r(z)$ one can write
\begin{equation}
W'[r(z+1)]=s^2 r''(z)+2W'[r(z)]-W'[r(z-1)].
\label{24}
\end{equation}
As shown in \cite{gfjadw94}, in such 
a lattice localized bell-shaped travelling-waves solutions
can exist if $W(r)$ has a hard anharmonicity in the region
$r<0$. Note that the acoustic solitons correspond to a 
localized contraction
of the chain and therefore the function $r(z)$ should be completely
negative.

It is evident by following the above line of argumentation for 
Klein-Gordon chains, that the pulse $r(z)$ has to be symmetric
and must have only one maximum. Any deviation from this leads to
a non-uniqueness in the definition of the potential. This implies
that acoustic chains admit at the best moving kink solutions 
(non-topological) in
the original variables $u_n(t)$, and even these kink solutions
have to have reflection symmetry. 

Suppose the function $r(z)$ satisfies the conditions for the pulses
given in Subsec. IIIA. Then due to the symmetry of $r(z)$, 
\begin{equation}
W'(r_1)=W'(r_0)+\frac{1}{2}s^2 r''(0)
\label{25}
\end{equation}
where $r_0=r(0)~$, $r_1=r(1)$. In order to find the unknown 
function $W'(r)$ we
have to solve the initial value problem (\ref{24}) where the initial
condition should be a the function $W'(r)$ on the
interval $r_0<r<r_1$. If $W'(r)$ is defined on this
interval, using Eqs. (\ref{24}) and (\ref{25}), one can construct the 
function $W'(r)$ for $r>r_1$ which will be 
 uniquely defined (see Fig. 4) for each $W'(r)$. This means
that we can choose $W'(r)$ in $[r_0,r_1]$ arbitrarily.
\begin{figure}[htbp]
\centerline{
\psfig{figure=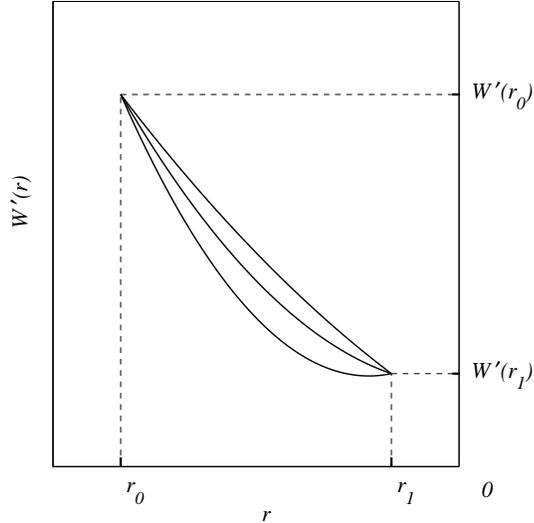,height=200pt,angle=0}
}
\caption[Fig. 4]
{ 
The schematic representation of the creation of the
potential $W(r)$ for acoustic chains. Different initial value choices
will generate different potentials for one and the same TW profile.
}
\label{fig4}
\end{figure}

Therefore we find that for given $\{ r(z),s \}$ a
countable infinite dimension of the space of solutions $W(r) \in C^2$ 
exists. 
Each function from this set supports one and the same $r(z)$ as
an exact solution with one and the same velocity.
However, the function constructed in this way from an
arbitrary initial value will be non-analytic in general 
(it is easy to show
that all functions will be twice differentiable).

To avoid the problem of generating non-analytic potential functions,
we found another way of constructing the potential.
Suppose $r(z)$ is a pulse. Then $W'[r(z)] \equiv \rho(z)$ is
also a pulse. Let us rewrite Eq. (\ref{24}) as
\begin{equation}
r''(z)=\frac{\rho(z+1)-2\rho(z)+\rho(z-1)}{s^2}~.
\label{26}
\end{equation}
Now instead of defining $r(z)$ we define $\rho(z)$ (symmetric
bell-shaped pulse). If this function is analytic, the rhs of
Eq. (\ref{26}) is also analytic. Then simply integrating this rhs
twice, we find an analytic function $r(z)$ which is also of
a bell-shaped form provided $\rho(z)$ decays for large $z$ faster
than $1/z^2$. 
Having $r$ as a function of $\rho$ (on a half axis $z$)
we can invert this dependence and consider $\rho$ as a function
of $r$. This gives us the force function $W'(r)$. Notice that
by that we can avoid generating non-analytic potentials.

Finally, let us take a look at the asymptotic behavior
of the solitary solution.
Suppose $\rho(z) \sim e^{-\mu z}$ for $z \rightarrow \infty$
and $W'(r) \simeq Cr$ for $r \rightarrow 0$.
Linearising Eq. (6), we obtain
\begin{equation}
\frac{s^2}{C}=2 \frac{\cosh \mu -1}{ \mu^2}>1~.
\end{equation}
Thus, each moving acoustic soliton is supersonic.

In the following example we will illustrate how to construct the interaction
potential from a given solution profile.
Suppose $\rho(z)=\rho_0~ \mbox{sech}^4 (\mu z)$. We substitute it
into the rhs of Eq. (\ref{26}) and integrate it
twice. Using the formula
\begin{eqnarray}
\nonumber
&&\int \int \mbox{sech}^4(\mu z) dz dz=-\frac{1}{6 \mu^2 \cosh^2 (\mu z)}\\
&&+\frac{2}{3 \mu^2} \ln [\cosh (\mu z)]+K_1 z + K_2~
\end{eqnarray}
we calculate the function $r(z)$. In order to satisfy the boundary
conditions, we put $K_1=0$, $K_2=0$.  As a result, we obtain
\begin{equation}
r(z)=\frac{\rho_0}{3 s^2 \mu^2} \left [ \ln \xi(z)+ 
\sqrt{\rho(z) \over\rho_0}  
\left(1-\frac{2(\sinh^2\mu-1)-\xi(z)}{\xi^2(z)} \right) \right ]~,
~\xi(z)=1+\sqrt{\frac{\rho(z)}{\rho_0}}\sinh^2\mu~.
\end{equation}
We cannot express the force function $W'(r)$ explicitly
from the above formula, but we have the inverse relation
$r(z)=D[\rho(z)]$ where D is a function inverse to $W'$.
The potential $W(r)$ can be calculated numerically.

%
%
\section{Breathers of the DNLS-type equations}

Here we study a general nonlinear chain governed by 
Eq. (\ref{3}). It is already known that these systems
have standing breather solutions (see, e.g., 
\cite{ma94}). The standing breather is defined
as a spatially localized solution which is periodic in time.
The general breather solution with frequency $\omega$, velocity
$s$ and wavenumber $q$ can be chosen to be
\begin{eqnarray}
&&\phi_n(t)=\Phi(n-st,\Omega t)
\;\;,\;\;
\nonumber \Phi(x\rightarrow \pm \infty ,y)\rightarrow 0\;\;,\;\;\\
&&\Phi(x,y)=\Phi(x,y+2\pi)\;\;.
\end{eqnarray}
Periodicity of $\Phi(x,y)$ in $y$ allows to expand it into a Fourier
series:
\begin{equation}
\Phi(x,y)=\sum_k \Phi_k(x) {\rm e}^{ik\Omega t} \;\;.
\end{equation}
For DNLS systems
these solutions may have only one nonzero Fourier harmonic w.r.t. time.
Since the DNLS equation has a gauge symmetry $\phi_n(t) 
\rightarrow e^{i\omega t} \psi_n(t)$, we can actually always
transform a breather solution into a stationary pulse solution
(note that this is not possible for breather solutions of
e.g., Klein-Gordon or acoustic chains). As a result Eq. (\ref{3})
can be rewritten as 
\begin{eqnarray}
i {\dot \psi}_n &+&\kappa \psi_n +C(\psi_{n+1}+\psi_{n-1})+
\nonumber
F(|\psi_n|^2)(\psi_{n-1}+\psi_{n+1})\\
&+& G(|\psi_n|^2)\psi_n=0~
\label{29b}
\end{eqnarray}
where $\kappa=-\omega-2C$.
Then we can define  $\psi_n(t) = R(z)+iI(z)~$, $z=n-st~$, 
where the functions $R$ and $I$ are real. Separating real and imaginary
parts of Eq. (\ref{29b}) we obtain the unknown nonlinear functions
$F$ and $G$ expressed in terms of breather envelope $R$ and $I$ and
breather parameters
\begin{eqnarray}
F[R^2(z)+I^2(z)]&=&-C+\frac{[R^2(z)+I^2(z)]'}{2\left (
R(z)[I(z-1)+I(z+1)]-I(z)[R(z-1)+R(z+1)]\right )}~,\\
\label{29d}
G[R^2(z)+I^2(z)]&=&-\kappa-\frac{s}{R(z)} \left [ I'(z)-
\frac{[R^2(z)+I^2(z)]'}{2\left (I(z)-
\label{29c}
\frac{I(z-1)+I(z+1)}{R(z-1)+R(z+1)}R(z)\right )} \right ]~.
\end{eqnarray}


We are interested in symmetric profiles, i.e., 
$u(z) \equiv \sqrt {R^2(z)+I^2(z)}=u(-z)$ in order to ensure 
single value properties  of the functions $F,G$.
It is easy see that the profile is symmetric and bell-shaped 
if, e.g.,  both $I(z)$ and $R(z)$ are symmetric or one
of these functions is symmetric and another one is
anti-symmetric. The lhs
of Eq. (\ref{29d}) is symmetric. If $R$ and $I$ are symmetric,
the functions $R(z-1)+R(z+1)$ and 
$I(z-1)+I(z+1)$ are also symmetric and the derivative of $R^2(z)+I^2(z)$
is anti-symmetric. Consequently, the rhs of this equation
is anti-symmetric. Obviously, this is possible only for a  trivial
solution $\psi_n(t)=0$. 

Clearly moving solutions seem to have more complicated internal
structures, as known from the solutions of the Ablowitz-Ladik system
\cite{al76}.  
These solutions can be represented in the following form:
\begin{equation}
\phi_n(t)=u_n(t)e^{i(qn-\omega t)}=u(z)e^{i(qn-\omega t)}~,~~
z=n-st~
\label{30}
\end{equation}
where $u(z)$ is a real envelope amplitude, $s$ is breather velocity,
$q$ its wave number and $\omega$  its frequency. Substituting 
this ansatz into Eq. (\ref{3}), we obtain two equations for
real and complex parts:
\begin{eqnarray}
(\omega -2C) u(z)&+&\cos {q} (C+F[u^2(z)]) [u(z-1)+u(z+1)]+
\label{31}
G[u^2(z)] u(z)=0~,\\
-s u'(z)&+& \sin {q} (C+F[u^2(z)]) [u(z+1)-u(z-1)]=0~.
\label{32}
\end{eqnarray}
After straightforward calculations the unknown nonlinear functions
$F$ and $G$ can be expressed in terms of $u$ and the system parameters:
\begin{eqnarray}
F[u^2(z)]&=&-C - \frac{s}{\sin {q}}\cdot \frac{u'(z)}{u(z+1)-u(z-1)}~,
\label{33}\\
G[u^2(z)]&=& -(\omega-2C)-\cos {q} (C+F[u^2(z)]) 
\nonumber     \frac{u(z-1)+u(z+1)}{u(z)}\\
         &=& -(\omega-2C)-\frac{s}{\tan {q}}\cdot \frac{u'(z)}{u(z)}
\label{34}   \cdot \frac{u(z-1)+u(z+1)}{u(z+1)-u(z-1)}.	 
\end{eqnarray}

If the envelope function $u(z)$ satisfies the conditions
\begin{itemize}
\item $u(z \rightarrow \pm \infty) \rightarrow 0 $ 
\item $u(-z)=u(z)$
\item $u(z)$ is monotonic in $[0,+\infty[$
\item $u(z)$ is analytic in $[0, +\infty[$,
\end{itemize}
we can postulate again (similarly to Sections III and IV) that 
for any envelope $u(z)$ defined as above and the set 
of parameters ($s,q,\omega$) one can uniquely define the 
nonlinearity for the equation given by functions $F$ and $G$.

\subsection{Examples}

Let us consider the particular case when $u(z)=u_0\mbox{sech}^2 (\mu z)$.
Substituting this expression into Eqs. (\ref{33}) and ({\ref{34}),
 we obtain functions $F$ and $G$:
\begin{eqnarray}
F(u^2)&=& -C + \frac{s\mu}{\sin q ~ \sinh 2\mu} 
\left [1+\frac{\sinh^2\mu }{u_0} u \right]^2~, \\
\nonumber
G(u^2)&=& -(\omega-2C)-\frac{2 s \mu}{\tan q ~ \sinh 2\mu} \\
&\times& \left (\cosh {2\mu} - \frac{\sinh^2 \mu}{u_0} u \right )~.  
\end{eqnarray}

\subsubsection{The Ablowitz-Ladik equation}

Let us look at the particular case when $G \equiv 0$.
In this case we have only one unknown nonlinear function,
$F$. After simplifying the ansatz (\ref{33})-(\ref{34}) we obtain
\begin{eqnarray}
\frac{u'(z)}{u(z)}&=&-
\label{37}
\frac{(2C-\omega)\tan {q}}{s}\cdot 
\frac{u(z+1)-u(z-1)}{u(z-1)+u(z+1)}~,\\
F[u^2(z)]&=&-C+\frac{(2C-\omega)u(z)}{\cos q [u(z-1)+u(z+1)]}.
\label{38}
\end{eqnarray}

In the particular case  $u(z)=u_0 \mbox{sech} (\mu z)$ we obtain
the quadratic function
\begin{equation}
F(u^2)=-C+\frac{2C-\omega}{2 \cos q ~ \cosh \mu} 
\left [1+\frac{\sinh^2 \mu}{u_0^2}u^2 \right ]~.
\end{equation}
We assume
\begin{equation}
\frac{2C-\omega}{2 \cos q ~ \cosh \mu}=C~,~~ 
\frac{2C-\omega}{2 \cos q ~ \cosh \mu}\cdot \frac{\sinh^2 \mu}{u_0^2}
=\frac{\lambda}{2}~,~~\lambda>0~.
\end{equation}
Eq. (\ref{37}) yields
\begin{equation}
2C-\omega=\frac{s\mu }{\tan q ~ \tanh \mu}~.
\end{equation}
We can rewrite these equations in more common way, expressing the
parameters of the solution $s, \omega$ and $u_0$ through $q$ and
$\mu$:
\begin{equation}
u_0=\sqrt{\frac{2C}{\lambda}}\sinh \mu~,
~\omega=2C [1-\cosh \mu ~ \cos q]~,~s=2C \frac{\sinh \mu}{\mu} \sin q~.
\end{equation}
This corresponds to the well-known integrable Ablowitz-Ladik
equation.

\subsubsection{DNLS with local nonlinearity}
Now let us look at a well-known equation of the DNLS
family. In the case of $F(u^2)\equiv 0$, we have
\begin{equation}
i \dot{\phi}_n +C(\phi_{n+1}-2\phi_n+\phi_{n-1})+
G\left (|\phi_n|^2 \right) \phi_n=0~.
\label{s3}
\end{equation}
We substitute the ansatz (\ref{30}) and consider  Eq. (\ref{31}) 
that for this particular case  takes the following form:
\begin{equation}
u'(z)=\alpha [u(z+1)-u(z-1)]~,~~\alpha=\frac{C\sin q}{s}~.
\label{41}
\end{equation}
The absence of a to be defined function in this equation, contrary
to the previous examples, makes this equation an equation for
the pulse shape. Let us show that a pulse shaped function $u(z)$
can not satisfy Eq. (\ref{41}). 
Suppose first that our solution $u(z)$ is periodic with some 
large period $L$. In this case we can expand the solution 
into Fourier series
\begin{equation}
u(z)= \sum_{m=-\infty}^{+\infty}u_m 
\exp {\left ( im\frac{2 \pi}{L} z\right )} ~.
\label{ds2}
\end{equation}
Substituting this expansion into Eq. (\ref{41}) we obtain the
algebraic equation
\begin{equation}
m \frac{2 \pi}{L}= 2 \alpha \sin \left (m \frac{2 \pi}{L} \right )
\label{ds1}
\end{equation}
where $m$ is unknown integer. The equation
$x=2\alpha \sin {x}$ always has always a finite
number of roots for any non-zero $\alpha $. 
Since $m$ is integer, we can actually solve (\ref{ds1}) only
for some specific values of $\alpha$. 
This does not depend on $L$, so we can consider the limit $L\rightarrow
\infty$. A pulse solution $u(z)$ would require an infinite number
of nonzero harmonics in Eq. (\ref{ds2}). 
Therefore it is impossible to satisfy Eq. (\ref{41}) with a pulse shaped
$u(z)$. Consequently, 
Eq. (\ref{s3}) does not admit moving breathers. 

It appears to be not possible to extend this method to 
systems like acoustic or KG chains. Note, that the spectrum 
of a DNLS-breather consists only of one frequency 
and therefore can be transformed into two differential
delay/advance equations while acoustic/KG breathers  will have an infinite
 number of harmonics and obviously cannot be rewritten as a
countable number of retarded and advanced ODE's.

\section{Continuation of moving solutions for conservative systems}

Let us discuss the question, whether a conservative system which allows
for a certain moving solution and has been generated by our
inverse method, has this solution as an isolated one, or as
a part of a smooth family of solutions. In other words, we consider a
given moving solution, generate the equations of motion, and
search in the phase space vicinity of our solution for other
moving solutions. That calls for a linear stability analysis
of the phase space flow around the given moving solution. 
Since our solution has some uniform asymptotic ground state
(we assume that the parameters are such that the uniform
asymptotic state is a ground state, see discussions above),
we know that a part of the linear stability analysis
spectrum will be given by just solving the eigenvalue problem
of linearized fluctuations around the ground state. The
eigenvectors will be plane waves ${\rm e}^{i(qn-\omega_qt)}$
(they will be deformed
in the center of our moving solution) and their spectrum 
is given by some dispersion relation $\omega_q$. 
Note that due to space discreteness $\omega_q$ is periodic
in $q$. Let us search
for $q$ values for which the plane wave can be cast into the form
\begin{equation}
{\rm e}^{i(qn-\omega_qt)} = h_q(n-st) \;\;.
\end{equation}
This is possible if 
\begin{equation}
qs=\omega_q\;\;.
\label{s10}
\end{equation}
Consider first the case of an acoustic chain. For small $q$
we have $\omega_q=vq$ with $v < s$ (because all moving solutions
will be supersonic, see discussions above). Consequently, 
there is only the trivial solution $q=0$ of Eq. (\ref{s10}), which
simply corresponds to a shift of the center of mass of the 
acoustic chain. We can always work in the frame where the center
of mass is resting at zero. Consequently, for acoustic chains
we do not find plane waves which move with the same velocity
as the original solution. That implies that there are no small
perturbations of our solution which have plane wave asymptotics
and yield again a moving solution. Thus, we conclude that if moving
solutions are coming in families, all solutions on the family will
be localized in space. 

Consider now the case of an optical chain (KG case). Since
$\omega_{q=0}\neq 0$, we will always find at least one nonzero
$q$-value (in general it will be a finite number of $q$ values
which tends to infinity as $s \rightarrow 0$) which 
solves Eq. (\ref{s10}).
In that case we do know that there will exist perturbations
with plane wave asymptotics which deform our original localized
moving solution into a partly delocalized one, which is now 
in addition characterized by a nonzero amplitude in the asymptotics
(i.e., the kink has oscillating tails at $|z| \rightarrow \infty$).
Some numerical results that confirm the above-mentioned 
arguments are given in \cite{sze99}.
 
To further analyse this, we performed a numerical continuation
of a moving solution using the pseudo-spectral method which
is essentially a Newton method in $q$ space (see, e.g.,  
\cite{jcerf90},\cite{yzjceavs97}). We chose the moving kink solution
(\ref{20}) which yields the potential (\ref{21}). The numerical
method traces the phase space of the system for moving solutions
nearby the starting one with slightly changed velocities and
shapes. 
\begin{figure}
\centerline{
\psfig{figure=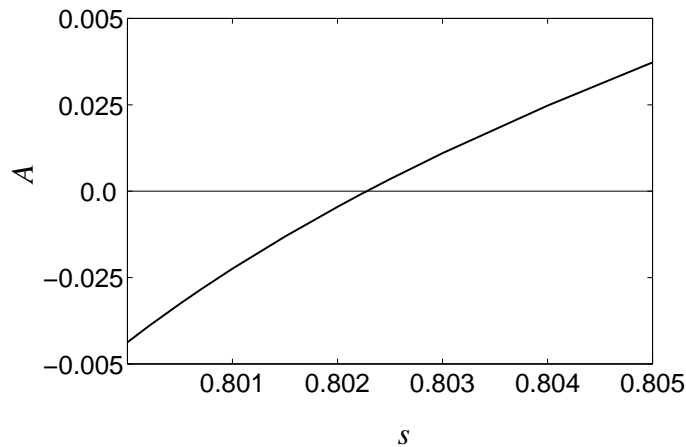,height=180pt,angle=0}
}
\caption[Fig. 5]
{
Dependence of the amplitude in the asymptotics of a moving kink
on the velocity for a given equation (see text).
}
\label{fig5}
\end{figure}
In Fig. 5 we show the dependence of the amplitude $A$
of the asymptotics versus velocity $s$. We indeed find that
our chosen solution (\ref{20}) (with $C=1$, $\mu=\ln {(1+\sqrt{2})}$,
$s=1/[\sqrt{2}\ln {(1+\sqrt{2})}]\simeq 0.802278$ in this 
particular case) can be continued, but it thus
gets dressed with plane wave asymptotics. The change of sign in
the amplitude implies a phase change from $0$ to $\pi$. 
A slightly changed
potential will exhibit similar solutions but with slightly shifted
curves in Fig. 5 . Thus, it follows that the moving solution with
uniform asymptotics is structurally stable, i.e., has a similar
solution with uniform asymptotics for slightly changed equations of motion. 
This follows from the fact that the crossing of the $A(s)$ curve in
Fig. 5 with $A=0$ is a generic intersection.

\section{Moving pulses in higher lattice dimensions}

So far we have been discussing moving pulses and kinks in
one-dimensional lattices. In this chapter we will show that
the inverse method can be easily generalized to higher space
dimensions for reaction-diffusion equations, provided we take
into account two or more components. 

Let us consider a two-dimensional lattice. The function $u_n(t)$
will now depend on two lattice indices $u_{m,n}(t)$. The differences
$(u_{n-1}+u_{n+1}- 2u_n)$ will now turn into some general discrete
Laplacians $D(u_{m,n})$. Assuming a moving solution in the form
\begin{equation}
u_{m,n}(t)=u(m-s_xt,n-s_yt)\equiv u(x,y)
\end{equation}
we arrive at the equations
\begin{equation}
-s_xu_{,x} - s_yu_{,y} + D[u(x,y)] = f(u)\;\;.
\end{equation}
Fixing a value of $u$ we obtain a line in the $\{x,y\}$ space and
since the rhs should not change, the lhs should be constant on this
line - a very restrictive condition. If we instead consider
two components moving in the same directions with same velocities,
the equations become
\begin{eqnarray}
-s_xu_{,x} - s_yu_{,y} + D[u(x,y)] = f(u,v)\;\;, 
\label{s40-1}\\
-s_xv_{,x} - s_yv_{,y} + D[v(x,y)] = g(u,v)\;\;.
\label{s40-2}
\end{eqnarray}
Again fixing a value for $u$ we obtain some line in $\{x,y\}$.
If we consider a functions $u,v$ decaying to zero at infinities,
this line will be a closed loop.
Let us assume that $v$ is not constant on the  loops of constant $u$. 
That helps,
but still if we fix some point on the loop with some
given value of $v$ there will be a countable number of 
other points $p_i$ on the loop 
where $v$ takes the same value. Then the lhs's of Eqs.
(\ref{s40-1}) and (\ref{s40-2}) have to be equal in these points.
We can satisfy this condition by demanding two symmetries but
only in the case when we have only {\it two} points $p_i$. First our pulse
functions $u,v$ should be invariant under reflections at a line
parallel to the direction of motion. This ensures that the first order
derivatives on the lhs's of Eqs. (\ref{s40-1}) and (\ref{s40-2}) will
be the same in all $p_i$. To ensure invariance of the Laplacians
in $p_i$ we only have to demand that the chosen direction of motion
(defined by $s_x,s_y$) is parallel to a reflection symmetry line
of the lattice. For instance, for a square lattice these will be only
the major lattice axes and the diagonals.
The initially assumed condition that $v$ varies along
the loop of constant $u$ can be, e.g., easily satisfied by considering
pulses which are symmetric under point reflections and whose symmetry
centers are shifted along the line of motion.
Note that contrary to the one-dimensional case the inverse method
yields the functions $f,g$ in a two-dimensional part of their phase
space $\{u,v\}$.

What if we add a third component? The conditions weaken again,
similarly to the case of two components in the one-dimensional lattice.
For instance, one can design pulses 
where two components are invariant
under point reflections with centers shifted along the line of motion,
and the third component will be off-centered from the line connecting
the two first centers. 

Let us consider a three-dimensional lattice and two components.
Fixing $u$ we now obtain a closed surface in $\{x,y,z\}$. Requiring
$v$ to generally vary on this surface, we find that $v$ will stay
constant at least on loops embedded on the surface. 
Since the lattice is invariant only under discrete symmetries,
we can not satisfy invariance of the 
lhs's of equations similar to Eqs. (\ref{s40-1}) and (\ref{s40-2})
 on this 
loop. Consequently, there exist no moving two-component pulses
in a three-dimensional lattice (and straightforwardly 
in any higher-dimensional
lattice). This is in contrast to the space-continuous case, where
space is invariant under continuous symmetries. Then we can
satisfy the invariance of the lhs's along the loop trivially
if both pulses are invariant
under rotations around a line pointing in the direction of their motion.
The initial condition that $v$ is constant only on loops (not on the
whole surface) is easily obtained by considering pulses $u,v$ with
shifted centers, just as in the two-dimensional case. 

Adding a third component $w$ to the three-dimensional lattice case 
reduces the problem of constant $v$
and $w$ on a loop to that in a countable set of points $p_i$.
Still we need a symmetry to ensure invariance of lhs's in the points
$p_i$. This is easily achieved in the case when we have only two 
points $p_i$ by demanding two symmetries.
First we need all three components
$u,v,w$ to be invariant under reflection at a plane which contains
the direction of motion. Secondly this plane has to be parallel
to any mirror reflection symmetry plane of the lattice.
Again the initial condition of having just countable sets of points
$p_i$ with coinciding values $\{u,v,w \}$ can be achieved by shifting
the centers of the three pulse components apart while staying on
one line - the direction of motion. 
For instance, for a cubic lattice with lattice points at $x=l,y=m,z=n$ and
$l,m,n$ integer reflection symmetry planes are $\{x,y,0 \}$, 
$\{0,y,z\}$, $\{x,0,z\}$, $\{x,\pm x,z\}$, $\{x,\pm z, z\}$, $\{x,y,\pm x\}$
among possible others. Any vector embedded in these planes is an allowed
moving direction.

%
%
\section{Conclusions}

In this paper we have studied several types of nonlinear lattice
systems. Contrary to most of papers on nonlinear lattices
where authors try to find a solution (either analytically or
numerically) of the given system, we approach the problem from
the opposite side - we look for the system (in fact, for the
interaction and/or on-site potentials) which admits some
specific solution. 
 
We have studied kinks and pulses in the Klein-Gordon
system, acoustic solitons in chains with nonlinear
intersite interactions and discrete breathers in the
nonlinear Schr\"{o}dinger-type systems. In all these cases
the method enables us to generate a unique on-site 
or interaction potential for a given pulse
or kink and its velocity if this solution satisfies
certain conditions.  
As a particular result, we have shown 
that the acoustic solitons are always supersonic. We also
conclude that nonzero Peierls-Nabarro barrier does not
prevent discrete kinks from propagating with constant
velocities.
In the case of discrete moving
breathers in  DNLS-type systems we create nonlinear
terms in the equation (\ref{3}) for given envelope
profile and breather frequency, wave number and velocity.

Our method is equally well suited for dissipative systems.
Systems
of coupled reaction-diffusion equations
do not possess one important property which 
is time reversibility and therefore despite being closely
related to the Klein-Gordon type equations, do not have 
pulse travelling-wave solutions in the one-dimensional case
for one component. We generalize the search for pulses to
higher lattice dimensions and find that moving pulses can be
easily obtained provided we also increase the number of components.

All presented results can be easily extended to systems
with longer range interactions, and to space continuous
systems (i.e., to partial differential equations). Note 
that the continuum limit of the considered difference equations
is easily recovered by choosing solitary wave profiles which
vary slowly along the lattice.

\acknowledgements

It is a pleasure to thank M. B\"{a}r, M. Or-Guil and A. A. Ovchinnikov 
for helpful discussions.










\end{document}